# Mössbauer Characterization of an Unusual High-spin Side-on Peroxo-Fe$^{3+}$ Species in the Active Site of Superoxide Reductase from *Desulfoarculus baarsii*. Density Functional Calculations on Related Models[‡]

Olivier Horner,*,[¶] Jean-Marie Mouesca,[#] Jean-Louis Oddou,[¶] Claudine Jeandey,[¶] Vincent Nivière,[§] Tony A. Mattioli,[±] Christelle Mathé,[±,§] Marc Fontecave,[§] Pascale Maldivi,[&] Pierre Bonville,[||] Jason A. Halfen,[⊥] and Jean-Marc Latour*,[¶]

[¶]*Laboratoire de Physicochimie des Métaux en Biologie, UMR CEA/CNRS/Université Joseph Fourier 5155, Département Réponse et Dynamique Cellulaires, CEA/Grenoble, 17 rue des Martyrs, 38054 Grenoble Cedex 9, France,* [#]*Laboratoire de Spectroscopie des Systèmes Biologiques, Service de Chimie Inorganique et Biologique, Département de Recherche Fondamentale sur la Matière Condensée, 17 rue des Martyrs, CEA/Grenoble, 38054 Grenoble Cedex 9, France,* [§]*Laboratoire de Chimie et Biochimie des Centres Redox Biologiques, UMR CEA/CNRS/Université Joseph Fourier 5047, Département Réponse et Dynamique Cellulaires, 17 rue des Martyrs, CEA/Grenoble, 38054 Grenoble Cedex 9, France,* [±]*Laboratoire de Biophysique du Stress Oxydant, Service de Bioénergétique, Département de Biologie Joliot-Curie, URA CNRS 2096, CEA/Saclay, 91191 Gif-sur-Yvette Cedex, France,* [&]*Laboratoire de Reconnaissance Ionique, Service de Chimie Inorganique et Biologique, Département de Recherche Fondamentale sur la Matière Condensée, 17 rue des Martyrs, CEA/Grenoble, 38054 Grenoble Cedex 9, France,* [||]*Service de Physique de l'Etat Condensé, Département de Recherches sur l'Etat Condensé, les Atomes et les Molécules, CEA/Saclay, 91191 Gif-sur-Yvette Cedex, France,* [⊥]*Department of Chemistry, University of Wisconsin-Eau Claire, 105 Garfield Avenue, Eau Claire, Wisconsin 54702 USA.*

[‡]This work was supported in part by the Regional Council of the Ile-de-France (S.E.S.A.M.E. equipment grant to T.A.M.) and the National Science Foundation grant (CHE-0078746 to J.A.H.).

# LIST OF ABBREVIATIONS[1]

---

[1] Cys : cysteine.

DFT : Density Functional Theory.

Dfx : the enzyme from the sulfate-reducing bacterium *Desulfovibrio desulfuricans*.

EDTA : Ethylene Diamine Tetraacetate.

EFG : Electric Field Gradient.

EPR : Electron Paramagnetic Resonance.

EXAFS : Extended X-ray Absorption Fine Structure

$[Fe^{II}(S^{Me2}N_4(tren))]^+$ : this complex is prepared by combining 2 equivalents of 3-methyl-3-mercapto-2-butanone with $FeCl_2$ in methanol, adding 1 equivalent of tren.

HOMO : Highest Occupied Molecular Orbital

ICSD : Inorganic Crystal Structure Database.

LCAO : Linear Combination of Atomic Orbitals

$L^8py_2$ : 5-bis-(2-pyridylmethyl)-1,5-diazacyclooctane.

LUMO : Lowest Unoccupied Molecular Orbital

MCD : Magnetic Circular Dichroism

Nlr : the enzyme from the archae *Pyrococcus furiosus*.

$N_4Py$ : *N,N*-bis(2-pyridylmethyl)-*N*-bis(2-pyridyl)methylamine)

PDB : Protein Data Bank.

Py : pyridine

RR : Resonance Raman.

SOR : Superoxide reductase.

SQUID : Superconducting Quantum Interference Device

TpivPP : meso-α,α,α,α-tetrakis(*o*-pivalamidophenyl)porphyrin

Tris : tris(hydroxymethyl)aminomethane

tren : tris(1-aminoethyl)amine.

ZFS : Zero Field Splitting

# ABSTRACT

Superoxide reductase (SOR) is an Fe protein that catalyzes the reduction of superoxide to give $H_2O_2$. Recently, the mutation of the Glu47 residue into alanine (E47A) in the active site of SOR from *Desulfoarculus baarsii* has allowed the stabilization of an iron-peroxo species when quickly reacted with $H_2O_2$ (Mathé, C.; Mattioli, T. A.; Horner, O.; Lombard, M. ; Latour, J.-M.; Fontecave, M.; Nivière, V. (2002) Identification of iron(III) peroxo species in the active site of the superoxide reductase SOR from *Desulfoarculus baarsii*, *J. Am. Chem. Soc. 124*, 4966-4967). In order to further investigate this non-heme peroxo-iron species, we have carried out a Mössbauer study of the $^{57}$Fe enriched E47A SOR from *D. baarsii* reacted quickly with $H_2O_2$. Considering the Mössbauer data, we conclude, in conjunction with the other spectroscopic data available and with the results of density functional calculations on related models, that this species corresponds to a high-spin side-on peroxo-$Fe^{3+}$ complex. This is one of the first example of such a species in a biological system for which Mössbauer parameters are now available : $\delta_{/Fe}$ = 0.54 (1) mm/s, $\Delta E_Q$ = - 0.80 (5) mm/s and the asymmetry parameter $\eta$ = 0.60 (5) mm/s. The Mössbauer and spin-Hamiltonian parameters have been evaluated on a model from the side-on peroxo complex (model **2**) issued from the oxidized iron center in SOR from *Pyrococcus furiosus*, for which structural data are available in the literature (Yeh, A. P., Yonglin, Y. , Jenney, Jr., F. E., Adams, M. W. W., Rees, D. C. (2000) Structures of the superoxide reductase from *Pyrococcus furiosus* in the oxidized and reduced states, *Biochemistry 39*, 2499-2508). For comparison, similar calculations have been carried out on a model derived from **2** (model **3**), where the $[CH_3-S]^{1-}$ group has been replaced by the neutral $[NH_3]^0$ group (Neese, F., Solomon, E. I. (1998) Detailed spectroscopic and theoretical studies on $[Fe(EDTA)(O_2)]^{3-}$: electronic structure of the side-on ferric-peroxide bond and its relevance to reactivity, *J. Am. Chem. Soc. 120*, 12829-12848). Both models **2** and **3** contain a formally high-spin $Fe^{3+}$ ion (*ie* with empty minority spin orbitals). We found however a significant fraction (~ 0.6 for **2**, ~ 0.8 for **3**) of spin (equivalently charge) spread over two occupied (minority spin) orbitals. The quadrupole splitting value for **2** is found negative and match quite well the experimental value. The computed quadrupole tensors are rhombic in the case of **2**, and axial in the case of **3**. This difference originates directly from the presence of the thiolate ligand in **2**. A correlation between experimental isomer shifts for $Fe^{3+}$ mononuclear complexes with computed electron densities at the iron nucleus has been built and used to evaluate the isomer shift values for **2** and **3** (0.56 mm/s and 0.63 mm/s, respectively). A significant increase of isomer shift value is found upon going from a methylthiolate to a nitrogen ligand for the $Fe^{3+}$ ion, consistent with covalency effects due to the presence of the axial thiolate ligand. Considering that the isomer shift value for **3** is likely to be in the 0.61 – 0.65 mm/s range (Horner, O., Jeandey, C.; Oddou, J.-L.; Bonville, P. ; McKenzie, C. J., Latour, J.-M. (2002) Hydrogenperoxo [(bztpen)Fe(OOH)]$^{2+}$ and its deprotonation product peroxo [(bztpen)Fe(O$_2$)]$^{+}$ studied by EPR and Mössbauer spectroscopies. Implication on the electronic structure of peroxo model complexes, *Eur. J. Inorg. Chem.*, 3278-3283. ), the isomer shift value for a high-spin $\eta^2$-O$_2$ $Fe^{3+}$ complex with an axial thiolate group can be estimated to be in the 0.54 – 0.58 mm/s range. The occurrence of a side-on peroxo intermediate in SOR is discussed in relation to the recent data published for a side-on peroxo-$Fe^{3+}$ species in another biological system (Karlsson, A.; Parales, J. V.; Parales, R. E.; Gibson, D. T.; Eklund, H.; Ramaswamy, S. (2003) Crystal structure of naphthalene dioxygenase: side-on binding of dioxygen to iron, *Science 299*, 1039-1042).

Superoxide Reductase (SOR) is an iron enzyme that catalyzes the one electron reduction of superoxide $O_2^{-\circ}$ to give $H_2O_2$ according to the reaction :

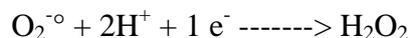

$$O_2^{-\circ} + 2H^+ + 1\ e^- \dashrightarrow H_2O_2$$

and which is involved in the mechanism of oxygen detoxification in anaerobic and microaerophilic microorganisms (*1,2*). The enzymes from the sulfate-reducing bacterium *Desulfovibrio desulfuricans* (Dfx) and that from the archae *Pyrococcus furiosus* (Nlr) have been structurally characterized (*3,4*). Their active site comprises in the reduced state an atypical $[Fe^{2+}(N-His)_4(S-Cys)]$ site. The penta-coordinated $Fe^{2+}$ ion is in a square pyramidal environment constituted by four equatorial histidine and an apical cysteinate ligands. It therefore possesses a vacant axial coordination site available to bind superoxide. In the oxidized state, this site is occupied by a glutamate residue which binds as a sixth ligand (*4,5*). The *D. desulfuricans, Desulfovibrio baarsii* and *Desulfovibrio vulgaris* enzymes possess an additional iron site (center I, as opposed to the active site called center II) which consists of a mononuclear $Fe^{3+}$ ion coordinated by four cysteinate residues in a distorted rubredoxin-type environment and is separated by *ca* 20 Å from center II (*3*). This center I is absent in the SORs from *P. furiosus* and *Treponema pallidum* and its biological role is still unknown (*2,4,6*).

Spectroscopic techniques (EPR, MCD, EXAFS) have been used extensively to characterize the SOR iron centers (*7*). In particular, the oxidized centers I and II have been widely studied by EPR spectroscopy. Oxidized centers II possess similar UV-visible and redox properties. Nevertheless, their EPR properties differ. Indeed, the enzymes which possess a center I (*eg* SOR from *D. desulfuricans*) exhibit a rhombic EPR signal (*8*) while this EPR signal is axial in the enzymes which lack center I (*eg* SOR from *T. pallidum* or *P. furiosus*) (*7*). Mössbauer spectroscopy was also used to characterize the as-isolated and oxidized SOR from *D. desulfuricans* (*8,9*). Indeed, Mössbauer parameters (isomer shifts and quadrupole splittings) have been determined for oxidized center I and center II (high-spin $Fe^{3+}$ ions), and reduced center II (high-spin $Fe^{2+}$ ion).

The reaction of the SORs from different origins with superoxide has been studied by pulse radiolysis methods (*10-14*). It is now generally assumed that the reaction of $O_2^{-\circ}$ with the $Fe^{2+}$ ion of reduced center II proceeds through an inner-sphere mechanism. The first step of the reaction is an extremely fast bi-molecular reaction of SOR with superoxide in a nearly diffusion-controlled process (~ $10^9\ M^{-1}\ s^{-1}$). In all enzymes studied, a first intermediate is formed that exhibits a broad absorption band at $\lambda \sim 600$ nm, ($\varepsilon \sim 3000 - 4700$ M$^{-1}$ cm$^{-1}$) and was proposed to be a peroxo-$Fe^{3+}$ species. However, depending on the enzymes studied, the following steps of the catalytic cycle differ. In the SORs from *Archaeglobus fulgidu and D. vulgaris*, the first intermediate becomes protonated to give directly the final products of the reaction, oxidized center II ($Fe^{3+}$) and $H_2O_2$ (*13,14*). In the SORs from *D. baarsii* and *T. pallidum*, an additional reaction intermediate has been identified, resulting from a single protonation process of the first intermediate, to give presumably an hydroperoxo-$Fe^{3+}$ species (*11,12,15*). This second intermediate is then transformed into the final product of the reaction $H_2O_2$. This occurs by a still uncharacterized protonation process with the help of the strictly conserved glutamate 47 (Glu47) residue which becomes a ligand of the $Fe^{3+}$ ion in the oxidized center II (*16*).

Mutation of Glu47 into alanine (E47A), which does not affect the binding of superoxide to the SOR from *D. baarsi* (*12, 15*), results in the stabilization of a peroxo-iron species when the enzyme is reacted with an excess of $H_2O_2$ (*16*). This species was shown to be in the $S = 5/2$ ground state by EPR spectroscopy. Trapping of this species by the E47A mutation permitted a RR characterization and the observed O-O and Fe-O vibrations were found to be consistent with a side-on peroxo-$Fe^{3+}$ species (*16*). Nevertheless, this conclusion was recently questioned from RR studies of model complexes (*17*) and DFT calculations which eventually advanced a hydroperoxo form as the most likely (*18*).

In order to further investigate this non-heme peroxo-iron species, we have carried out extensive Mössbauer studies of the $^{57}$Fe enriched E47A SOR from *D. baarsii* reacted with $H_2O_2$. Indeed, recent Mössbauer studies of peroxo-iron complexes have shown that the side-on peroxo and hydroperoxo forms exhibit different spectroscopic signatures owing to their different spin states and geometries (*17,19*). The value of the Mössbauer isomer shift deduced in the present study for the purported side-on peroxo-$Fe^{3+}$ species departs slightly from that expected from model complexes. We reasoned that the unusual presence of a cysteinate iron ligand might be responsible for the observed value. This prompted us to investigate in detail the effect of thiolate ligation on the Mössbauer parameters by i) studying model complexes of reduced center II ($Fe^{2+}$) in SOR by Mössbauer spectroscopy, ii) evaluating the Mössbauer (isomer shift excluded) and spin-Hamiltonian parameters by DFT calculations for a structural model of oxidized center II in E47A SOR reacted with $H_2O_2$, iii) establishing a correlation between experimental isomer shifts for mononuclear ferric complexes and computed electron densities at the iron nucleus, in order to estimate the isomer shift value for the model of oxidized center II. From our Mössbauer data and the results of DFT calculations, we conclude that the peroxo-iron species in E47A SOR is a high-spin side-on peroxo-$Fe^{3+}$ species in agreement with the RR study (*16*). The occurrence of a $\eta^2$-$O_2$ $Fe^{3+}$ intermediate in SOR will be discussed in relation to the experimental data already published and the recent proposal for the catalytic mechanism.

## MATERIALS AND METHODS

*Preparation of $^{57}$Fe SOR E47A from D. baarsii Samples and Model Complexes.* The purified $^{57}$Fe SOR E47A sample was prepared as already described (*12*) except that 98 % $^{57}$Fe instead of natural Fe was used to complement the *Escherichia coli* DH5α pMLE47A culture medium. For isotopic enrichment, the $^{57}$FeCl$_3$.6H$_2$O starting compound was obtained by dissolving $^{57}$Fe$_2$O$_3$ (AMT Ltd.) in 20 equivalents of concentrated HCl (Carlo Erba) under reflux and then by evaporating to dryness.

Three $Fe^{2+}$ complexes of the $L^8py_2$ ligand of the general formula $[L^8py_2Fe^{II}(X)]^+$ (X = SC$_6$H$_4$-*m*-CH$_3$ (**4**), SC$_6$H$_{11}$ (**5**) and CH$_3$CO$_2$ (**6**) were prepared according to the literature procedure (Scheme 1) (*20*).

((Scheme 1))

*Spectroscopic Methods.* All Mössbauer measurements were performed as already described (*19*). One home-made sample holder able to generate an external magnetic field of 50 mT applied parallel to the Mössbauer γ-beam was also used (*21*). The samples for Mössbauer spectroscopy contained *ca* 1.5 mM or 4 mM $^{57}$Fe in a 200 μL nylon cell. The analysis of the Mössbauer data was made as already described (*19*). In order to determine an accurate value of the isomer shifts, a special procedure was followed when fitting the spectra. Indeed, the contribution of the oxidized center I in SOR from *D. baarsii* was first at all subtracted from the experimental data. This difference spectrum was then fitted by considering only the contribution of the oxidized center II in SOR from *D. baarsii*. This theoretical spectrum was then subtracted from the experimental spectrum, and this new difference spectrum was further fitted by considering only the contribution of oxidized center I. This procedure was repeated until a convergence of the Mössbauer parameters was observed (7 or 8 cycles), which allowed an accurate determination of the isomer shift values. The Euler angles α, β and γ define the orientation of the [A] tensor relative to the EFG tensor (the [A] and [g] tensors are assumed here to be colinear).

Magnetization measurements were performed with a SHE SQUID 700 magnetometer operating at six magnetic fields in the range 0.5 T – 5 T over the temperature domain 5 K – 200 K. After concentration in a deuterated buffer, the sample (5.1 mM) was de-aerated under argon and 115 µL was transferred into a quartz sample bucket within a glovebox under argon. Upon immediate removal from the glovebox the sample (under argon in a small container) was frozen in liquid nitrogen and introduced in the magnetometer. The magnetization of the protein was obtained by subtraction of the buffer magnetization measured in the same conditions according to the general procedure outlined by Day *(22)*. Simultaneous fitting of the six isofield curves was performed with a home-made FORTRAN program *(23)*. EPR spectra were recorded as already described *(19)*. Resonance Raman spectra were recorded and analyzed as already described, with a laser excitation at 647.1 nm and at 15 K *(16)*. The intermediate $Fe^{3+}$-peroxo species was prepared by mixing 6 eq. of $H_2O_2$ with ferrous SOR and rapidly freezing in liquid nitrogen within 3 seconds of mixing. The same $Fe^{3+}$-peroxo could be prepared (as determined by RR spectroscopy) by mixing 6 eq. of $H_2O_2$ to ferric SOR which was previously oxidized with 3 eq. of $K_2IrCl_6$ and then washed to remove excess $K_2IrCl_6$ (Mathé, Nivière, Mattioli, unpublished results). Possible Fenton chemistry and •OH-related protein damage at the $Fe^{3+}$ sites (as monitored by UV-visible absorption and RR spectroscopies) were not observed on the time-scale of the $Fe^{3+}$-peroxo species formation. It is known that treatment of some $Fe^{2+}$ complexes with $H_2O_2$ results in the formation of metastable $Fe^{3+}$-peroxo species (52); the mechanism is not completely understood but has been addressed experimentally (55,56). Furthermore, Tris buffer is a good •OH scavenger (57) which further minimizes the risks of protein damage.

*Computational Methods.* All calculations have been performed with the Amsterdam LCAO Density-Functional Programs (ADF 2.3) developed by Baerends *et al* *(24-28)*. We considered there only the potential refered to as «VBP» (Vosko, Wilk and Nusair's exchange and correlation energy *(30,31)* completed by non local gradient corrections to the exchange by Becke *(32)* as well as to the correlation by Perdew *(33)*. We used triple-zeta (plus polarization) basis sets for for all atoms.

*Models Used and Choice of Geometries.* Figure 1 shows the models used in this study :

((Figure 1))

The crystal structure of $Fe^{3+}$ SOR from *P. furiosus* has been determined by X-ray crystallography at 1.70 Å resolution (PDB entry 1DQI) (*4*). The coordinates are therefore readily available in order to build structural models suitable for Density Functional LCAO calculations. Model **1** corresponds to a simplified structure of the oxidized center II issued from the oxidized iron center of SOR (*P. Furiosus*), where residues His 16, His 41, His 47 and His 114 of the equatorial plane have been replaced by four neutral imidazole rings $[N_2C_3H_4]^0$, and where the axial residues Cys 111 and Glu 14 have been replaced by the charged $[CH_3-S]^{1-}$ and $[CH_3-CO_2]^{1-}$ groups, respectively (Table S6). In such a model, the **z** axis has been set along the Fe-S(Cys) bond. The **x** axis is then roughly defined along the Fe-N(His16,47) bonds and the **y** axis along the Fe-N(His41,114) bonds, with the projected S-C bond of the $[CH_3-S]^{1-}$ group along the (Fe-S) **z** axis being closer to the Fe-N **x** axis.

From model **1**, we then derived model **2**, obtained by replacing the $[CH_3-CO_2]^{1-}$ group (located along the **z** axis) by a side-on peroxo group $O_2^{2-}$ (Table S7). The O-O and Fe-O bond lengths have been set to the values used by Neese *et al* for high-spin $O_2^{2-}$ - $Fe^{3+}$ models that they built for their calculations on $[Fe(EDTA)(O_2)]^{3-}$ (1.41 and 2.05 Å, respectively) (*34*). These distances are consistent with those determined very recently for a side-on peroxo adduct crystallographically characterized for a naphthalene dioxygenase (*35*). Considering now the Fe-S bond, an EXAFS study of oxidized center II in SOR from *P. furiosus* yielded a bond length of 2.36 Å (*7*). Recently, Schearer *et al* structurally characterized two model complexes of the reduced and oxidized center II in SOR, $[Fe^{II}(S^{Me2}N_4(tren))]^{1+}$ and the hexacoordinate $[Fe^{III}(S^{Me2}N_4(tren))(CH_3CN)]^{2+}$ complexes (*36*). They measured the Fe-S bond lengths to be 2.31 and 2.33 Å, respectively. Upon reaction of $[Fe^{II}(S^{Me2}N_4(tren))]^{1+}$ with superoxide, an $Fe^{3+}$-hydroperoxo intermediate that possesses a Fe-S bond length of 2.33 Å, as determined by EXAFS, was isolated (*36*). Moreover, **4**, **5** and $[L^8py_2Fe^{II}(SC_6H_4\text{-}p\text{-}CH_3)]^{1+}$ exhibit an average Fe-S bond length of 2.30 Å (*20*). Therefore, we decided to set the initial value of the Fe-S bond length in **2** to 2.30 Å. We finally constructed model **3**, starting from model **2**, by further replacing the $[CH_3-S]^{1-}$ group by the neutral $[NH_3]^0$ group, with a Fe-$N(NH_3)$ bond length value of 2.10 Å (Table S8) (*34*).

Full geometry optimization was performed on models **2** and **3** (and **2'** and **3'** were imidazoles were replaced by $NH_3$). It is noteworthy that the optimized geometry for **2** matched very well that of model 5 from Kurtz *et al.* (*18*). These calculations allowed us to estimate the respective electron densities at the iron nucleus for all optimized structures which were converted to the respective isomer shifts (using the correlation of figure 7) 0.63 and 0.70 mm/s for **2** and **3** (0.68 and 0.74 mm/s for **2'** and **3'**). The value for **3** (and **3'**) is definitely out of the range observed for peroxoferric complexes of N donors which is confined to 0.60 – 0.65 mm/s. The reason for this discrepancy must be found in the constraints imposed by the protein backbone and the polydentate ligands used in biomimetic compounds. These observations led us to calculate the electronic structure for **2** and **3** by imposing the geometry of the metal and of the protein ligands to their positions deduced by X-ray crystallography in the oxidized SOR from *P. furiosus* as detailed above. The peroxo ligand was added at a chemically reasonable distance, as done by Neese *et al.* (*34*) in their calculation of the complex $[Fe(O_2)(edta)]^{3-}$. A geometry optimization of the angle $\theta$ between the (FeOO) plane and the Fe-N axis was carried out by rotating the O-O bond around **z** by steps of 15°. This led to the O-O bond being oriented along **y** for both **2** and **3**. The calculation of the electronic structure of **3** performed for this model led to an isomer shift value of 0.60 mm/s within the experimental range. Calculation of **2** gave an isomer shift value of 0.55 mm/s, again in agreement with the experimental value 0.54(1) mm/s. It is worth noting that all calculations point to a reduction of the isomer shift value of ca 0.08 - 0.10 mm/s when an axial amine ligand is replaced by a thiolate.

*Spectroscopic Parameters.* The EPR and Mössbauer parameters (isomer shift excluded, see below) were evaluated by first determining the electronic structures corresponding to models **2** and **3** using DFT electronic structures as provided by the ADF code. We then relied on a home-made code in order to compute the [g] and ZFS tensors [D] as well as the quadrupole tensor [Q]. In order to compute the [g] and ZFS tensors, we relied on the following expression :

$$g_{ij} \approx g_e \delta_{ij} - \frac{2\xi}{2S}\left(\sum_\alpha - \sum_\beta \right)\sum_n \frac{\langle o|L_i|n\rangle\langle n|L_j|o\rangle}{E_n - E_o} \qquad (4)$$

with $\{i,j\}=\{x,y,z\}$, the spin-orbit coupling constant $\xi = 404$ cm$^{-1}$ and S=5/2 for the d$^5$ Fe$^{3+}$ ion. The labels "$\alpha$" and "$\beta$" stand for the five filled majority and the five empty minority spins of the high-spin $Fe^{3+}$ ion, respectively. $|o\rangle$ stands for an occupied molecular orbital whereas $|n\rangle$ represents an empty one. The $E_n - E_0$ energy gaps, when promoting an electron from and to mainly Fe molecular orbitals (as would usually be the case for a $Fe^{2+}$ ion-containing complex within the minority spin set of orbitals) can be computed as Slater transition state energies ascribing half an electron to both $|o\rangle$ and $|n\rangle$. The difference in the corresponding half-occupied molecular orbital energies is then taken as a good estimate of the total energy difference $E_n-E_o$ between the two electronic structures $(o)^1(n)^0$ and $(o)^0(n)^1$. We will show below how this point is relevant for the oxidized (*ie* $Fe^{3+}$) models **2** and **3**.

The ZFS tensor [D] is obtained from Equation 4 by replacing $\xi$ with $-\xi^2$. The D and E ZFS parameters are then defined by :

$$\begin{cases} D = 3(D_{zz} - D_{iso})/2 \\ E = |D_{xx} - D_{yy}|/2 \end{cases} \qquad (5)$$

where $D_{iso} = \text{Tr}([D])/3$. We also computed the quadrupole tensor [Q] as (*37*) :

$$[Q] \approx \frac{1}{2}\left(e^2 Q <r^{-3}> (1-R_0)\right)\left(\sum_\alpha + \sum_\beta\right)\frac{1}{7}[\Omega] \qquad (6)$$

where $\Omega_{ij} = \langle\Phi|L_iL_j+L_jL_i-(2/3)\delta_{ij}L(L+1)|\Phi\rangle$. Q is the quadrupole moment and $(1-R_0)$ the Sternheimer factor. Numerically, $[Q] = 0.925[\Omega]$ mm.s$^{-1}$.

*Isomer Shift Correlation.* A number of small mononuclear $Fe^{3+}$ and biomolecules that cover the full range of $^{57}$Fe shifts were selected for calibrating the relationship between the experimental isomer shifts and the calculated electron densities at iron nuclei. The calibration process consisted of three steps. First, the structures of the complexes were obtained from the X-ray data available at the ICSD ([Fe(NO$_2$)(Py)(TpivPP)] and [Fe(CN)$_6$]$^{3-}$) or from some atomic coordinates already published elsewhere ([FeCl$_4$]$^{1-}$ and [Fe(EDTA)(O$_2$)]$^{3-}$, the latter corresponding to a computational model complex, see below). The structures of the biomolecules were constructed from the X-ray data available at the PDB, Oxidized center II in SOR and [Fe(SR)$_4$]$^{1-}$), after simplification of the biological residues ligated to the iron center ( R = -CH$_2$CH$_3$). Second, all previous structures were directly used for Density Functional LCAO calculations (see above) in order to compute electron densities at the $^{57}$Fe nuclei. Third, the calculated electron densities at iron nuclei were plotted *vs* the experimental isomer shifts, and the resulting graph was subjected to a linear regression analysis. All the experimental isomer shifts reported in this work are reported relative to an Fe metal standard at room temperature and refer to a sample temperature of 4.2 K.

## RESULTS AND ANALYSIS

The oxidation of E47A SOR from *D. baarsii* by $K_2IrCl_6$ or $H_2O_2$ is almost complete (see below) but a few percent of the unreacted enzyme containing reduced center II ($Fe^{2+}$) was still present (Scheme 2).

((Scheme 2))

Therefore, we decided to characterize the reduced center II both in the as-isolated and in the dithionite-reduced enzymes. Indeed, the spin-Hamiltonian parameters of reduced center II in SOR were extracted more easily in the latter case as detailed below.

*As-isolated and Dithionite-reduced E47A SOR from D. baarsii.* In the as-isolated E47A SOR from *D. baarsii*, center I contains a $Fe^{3+}$ ion, whereas center II contains a $Fe^{2+}$ ion (Scheme 2). The as-isolated E47A SOR from *D. baarsii* was investigated by magnetization measurements to quantify the ZFS parameters of the reduced center II ($Fe^{2+}$). Figure 2 shows the temperature dependence of the molar magnetization in the as-isolated enzyme with the $\beta B/kT$ ratio for different magnetic fields (isofield experiments) :

((Figure 2))

The data were simultaneously fitted within the spin-Hamiltonian formalism. Oxidized center I is characterized in EPR spectroscopy by resonances at $g_{eff}$ = 7.7, 5.7, 4.1 and 1.8 (*6,12,16*), which correspond to D > 0 and E/D ~ 0.08. These data are close to those obtained for the oxidized desulforedoxin from *Desulfovibrio gigas* (*38*). Indeed, oxidized center I has a distorted tetrahedral sulfur coordination very similar to that found in the desulforedoxin from *D. gigas* (*39*). The magnetic contribution of oxidized center I ($Fe^{3+}$) was estimated by fixing the D value in the fitting procedure to the value (D = 2.2 cm$^{-1}$) determined for the oxidized desulforedoxin from *D. gigas* (*38*). The best fit of the molar magnetization (solid line in Figure 2) gave for the reduced center II (high-spin $Fe^{2+}$ ion) the following parameters: g = 2.15, |D| = 5.2 cm$^{-1}$ and E/D = 0.28 ± 0.03.

Figure 3 shows the 4.2 K Mössbauer spectrum of the as-isolated E47A SOR from *D. baarsii* recorded with a magnetic field of 50 mT applied parallel to the γ–beam.

((Figure 3))

Two distinct spectral components are clearly distinguishable: a dominant central quadrupole doublet ((a) in Figure 3) and a magnetic spectral component extending from -5.0 to 5.0 mm/s ((b) in Figure 3). The quadrupole doublet accounts for 53 ± 3 % of the total absorption, whereas the magnetic component accounts for 47 ± 3 % of the total absorption. The Mössbauer parameters obtained for the quadrupole doublet ($\delta_{/Fe}$ = 1.06 (1) mm/s and $\Delta E_Q$ = 2.82 (3) mm/s) are characteristic of a high-spin $Fe^{2+}$ ion and correspond to the reduced center II of SOR from *D. baarsii*. These parameters are comparable to those published for the reduced center II of wild-type SOR from *D. desulfuricans* ($\delta_{/Fe}$ = 1.04 mm/s and $\Delta E_Q$ = 2.80 mm/s) (*9*). The magnetic component corresponds to the oxidized center I and has been fitted with a unique set of parameters shown in Table 1, the D and E/D values being fixed in the fitting procedure to 2.2 cm$^{-1}$ and 0.08, respectively (see above) :

((Table 1))

The isomer shift and quadrupole splitting values obtained for the magnetic spectral component ($\delta_{/Fe}$ = 0.29 (2) mm/s and $\Delta E_Q$ = - 0.79 (4) mm/s) are characteristic of the high-spin $Fe^{3+}$ ion in center I and compare well with the parameters obtained for the oxidized desulforedoxin from *D. gigas* ($\delta_{/Fe}$ = 0.25 (6) mm/s and $\Delta E_Q$ = - 0.75 (5) mm/s) (*38*).

Reduction of E47A SOR from *D. baarsii* with 1.1 equivalent of sodium dithionite occurs with immediate loss of the red color of the sample. In this case, both center I and center II contain a $Fe^{2+}$ ion (Scheme 2). The zero-field Mössbauer spectrum of the dithionite-reduced E47A SOR from *D. baarsii* consists of two quadrupole doublets (Figures S1, A) that have been fitted with the parameters in Table 1. Moreover, the spectrum of the dithionite-reduced enzyme was measured at 200 K in a parallel field of 7.0 T, and the reversed patterns of triplet and doublet structures show that $\Delta E_Q$ is positive and that η is less than 0.5 for both reduced sites (Figure S1, B) (*38,40*). Mössbauer spectra of the dithionite-reduced enzyme were also recorded at 4.2 K in variable fields (Figure S2). The Mössbauer data were fitted simultaneously within the spin-Hamiltonian formalism using the set of parameters shown in Table 1. In the fitting procedure, the ZFS parameters for reduced center I were fixed to the values published for reduced desulforedoxin from *D. gigas* (*38*).

*Oxidation of E47A SOR from D. baarsii by $K_2IrCl_6$.* In the oxidized enzyme, both center I and center II contain a $Fe^{3+}$ ion (Scheme 2). Indeed, center II in the E47A SOR from *D. baarsii* can be oxidized by a slight excess of $K_2IrCl_6$. The 4.2 K EPR spectrum of the oxidized enzyme consists of a large rhombic derivative signal at g = 4.3, which is characteristic of a high-spin $Fe^{3+}$ ion (oxidized center II) in a rhombic ligand field (*8,16*). The Mössbauer spectra of the E47A SOR from *D. baarsii* oxidized with $K_2IrCl_6$ recorded at 4.2 K in various magnetic fields applied parallel to the γ-rays are shown in Figure 4 :

((Figure 4))

43 ± 2 % of the experimental spectra obtained for the oxidized enzyme are accounted for by the contribution of the $Fe^{3+}$ ion from center I ((a) in Figure 4). In addition, 4 ± 2 % of the spectra ((c) in Figure 4) are accounted for by the contribution of $Fe^{2+}$ ion from center II showing that the oxidation is essentially quantitative. The major part of the spectra (53 ± 2 %) corresponds to the oxidized center II ((b) in Figure 4). It was satisfactorily fitted with Mössbauer and spin-Hamiltonian parameters ($\delta_{/Fe}$ = 0.47 (1) mm/s and $\Delta E_Q$ = - 0.53 (5) mm/s with η = 0.00 (5)) close to those published for the oxidized center II (gray form) in the wild-type SOR from *D. desulfuricans* ($\delta_{/Fe}$ = 0.50 (2) mm/s and $\Delta E_Q$ = - 0.53 (3) mm/s with η = 0.14) (*8*). This similarity suggests that the $Fe^{3+}$ ion in the oxidized center II of the mutated enzyme (for which no Glu47 ligand is available) is likely to be hexa-coordinated. Moreover, the $Fe^{3+}$ environments of the oxidized center II in both enzymes should be similar making water a plausible sixth ligand in the mutated enzyme. Indeed, in the zero-field Mössbauer study (T =77 K) of some dinuclear $Fe^{3+}$ complexes with some N/O ligands, the following Mössbauer parameters were obtained : $\delta_{/Fe}$ = 0.55 (1) mm/s and $\Delta E_Q$ = 0.60 (1) mm/s for a hexa-coordinated $Fe^{3+}$-OH center, and $\delta_{/Fe}$ = 1.51 (1) mm/s and $\Delta E_Q$ = 0.60 (1) mm/s for a hexa-coordinated $Fe^{3+}$-$OH_2$ center (*41*). These latter parameters (in particular when considering the quadrupole splitting value) are close to those determined for the E47A SOR from *D. baarsii* oxidized by $K_2IrCl_6$.

*Oxidation of E47A SOR from D. baarsii by $H_2O_2$.* In order to form peroxo-$Fe^{3+}$ species, the as-isolated $^{57}$Fe SOR E47A from *D. baarsii* was reacted with 6 equivalents of $H_2O_2$ at pH = 8.4 and immediately frozen within 5 seconds in liquid nitrogen, as already reported (*16*). An aliquot of this sample was analyzed by RR spectroscopy (Figure S3). The RR spectrum exhibits bands at 850 cm$^{-1}$ (no $^{57}$Fe isotopic shift) and 436 cm$^{-1}$ ($^{57}$Fe isotopic downshift of 2 cm$^{-1}$) which correspond to the ν(O-O) and ν(Fe-O) stretching modes of the iron-peroxo species, respectively, as previously reported (*16*). This control measurement clearly shows that the Mössbauer cell contains the peroxo-iron species under study. An aliquot of the Mössbauer sample was also studied in X-band EPR spectroscopy at 4.2 K and shows a large derivative signal at g = 4.3, which is characteristic of a high-spin $Fe^{3+}$ ion in a rhombic ligand field (data not shown) and is not due to oxidized center I (see above). Mössbauer spectra of the sample recorded in several magnetic fields applied perpendicular to the γ-rays, which allows a better separation of center I and oxidized center II contributions, are shown in Figure 5 :

((Figure 5))

The solid lines correspond to the best fit obtained with the parameters reported in Table 1. $Fe^{3+}$ ion from center I contributes for 52 ± 2 % of the total spectra and its contribution ((a) in Figure 5) was simulated with the Mössbauer and spin-Hamiltonian obtained for center I in the as-isolated enzyme (Table 1). 45 ± 2 % of the experimental spectra is accounted for by a second component with a large magnetic splitting ((b) in Figure 5) and which differ from the one observed in the spectra of the $K_2IrCl_6$-oxidized enzyme (see above). It can be assigned to a high-spin $Fe^{3+}$ site whose parameters are reported in Table 1. We assign this $Fe^{3+}$ ion site to the peroxo-iron species. The associated contributions in Figure 5 were fitted by allowing some anisotropy of the [A] tensor ($A_{av}/g_n\beta_n$ = -21 T). Moreover, these fits were improved by intoducing three Euler angles (α = 17° ; β = 80° ; γ = 60°) between the [A] and the EFG tensors. The isomer shift and quadrupole splitting values of the peroxo-iron species ($\delta_{/Fe}$ = 0.54 (1) mm/s and $\Delta E_Q$ = - 0.80 (5) mm/s, respectively, with η = 0.60 (5)) are clearly different from those obtained for center II oxidized by $K_2IrCl_6$, in particular concerning the isomer shift and the quadrupole splitting values ($\delta_{/Fe}$ = 0.47 (1) mm/s and $\Delta E_Q$ = - 0.53 (5) mm/s with η = 0.00 (5)). It must be noticed that the Mössbauer parameters of the peroxo-iron species are typical for high-spin $Fe^{3+}$ complexes and therefore are not unique for the peroxo complex. Finally, a minor contribution ((c) in Figure 5) is also discerned (3 ± 1 %) that corresponds to reduced center II of SOR E47A ($\delta_{/Fe}$ = 1.06 (2) mm/s and $\Delta E_Q$ = 2.82 (2) mm/s) and that has been simulated with the parameters of Table 1. It must be noted that this complex is remarkably stable in the E47A mutant since it is still present almost quantitatively within 5 seconds (90 ± 4 % of center II is in the peroxo-iron form).

*Evaluation of the Mössbauer and Spin-Hamiltonian Parameters for a $Fe^{3+}$-peroxo Species with an Axial Thiolate Ligand.* The set of Mössbauer and spin-Hamiltonian parameters that we have obtained for the peroxo-iron species of the E47A mutant of the SOR from *D. baarsii* match the properties reported for side-on peroxo-$Fe^{3+}$ model compounds (*eg* [($\eta^2$-$O_2$)Fe(EDTA)]$^{3-}$ or [($\eta^2$-$O_2$)Fe(N$_4$Py)]$^+$) except for its isomer shift value which is slightly lower : $\delta$ = 0.54 mm/s *vs* 0.61 - 0.65 mm/s for complexes with nitrogen ligands (*17,19,42,43*). Obviously, the SOR active site differs from these model compounds by the presence of an axial thiolate ligand. This ligand is expected to lower the isomer shift by covalency effects. However, to the best of our knowledge no high-spin $\eta^2$-$O_2$ $Fe^{3+}$ complexes mimicking the peculiar [N$_4$S] SOR environment have been characterized so far in the literature by Mössbauer spectroscopy. Therefore, it appeared important to us to quantify the effect of a thiolate ligand on the values of the Mössbauer parameters, and in particular on the isomer shift, when going from a (N/O) to a (N/O)/S side-on peroxo-iron species. We approached this question by combining experimental studies of model compounds and DFT calculations.

*(a) Model compounds.* First, we decided to obtain zero-field Mössbauer data (T = 4.2 K) for model compounds of reduced center II in SOR **4** ($\delta_{/Fe}$ = 0.93 (1) mm/s) and **5** ($\delta_{/Fe}$ = 0.93 (1) mm/s (Scheme 1 and Figure S4). For the sake of comparison, the zero-field Mössbauer spectrum of **6** (Scheme 1), where no axial thiolate ligand is present, has also been recorded ($\delta_{/Fe}$ = 1.06 (1) mm/s, Figure S4). It appears that replacement of the carboxylate ligand in **6** by a the thiolate ligand in **4** and **5** results in a decrease of the isomer shift value of ~ 0.13 mm/s (Table S5). This difference is explained by covalency effects associated to the thiolate ligand in complexes **4** and **5** (*44*). Taking these covalency effects into consideration for thiolato-bound peroxo-$Fe^{3+}$ species would lower the expected isomer shift value from the 0.61 – 0.65 mm/s range (*17,19,42,43*) to *ca* to 0.55 mm/s, indeed close to the experimentally determined value of the SOR peroxo-iron species.

*(b) DFT calculations.* We have evaluated by computational methods the Mössbauer (except for the isomer shift, see below) and the spin-Hamiltonian parameters of model **2**, a side-on $Fe^{3+}$-peroxo complex (S = 5/2) with a $[CH_3-S]^{1-}$ group in axial position (Figure 1 and Table S7). For comparative purpose, similar calculations have been performed on model **3** derived from **2** by replacing the $[CH_3-S]^{1-}$ group by the neutral $[NH_3]^0$ group (Figure 1 and Table S8). Let us start by some general considerations concerning the electronic structures computed for both models **2** and **3**, for which a description of the electronic structures is available in Figure 6 (the spectroscopic parameters – [g], ZFS and quadrupole tensors - computed in this paper are essentially determined by the minority spin orbitals, hence the choice of representing some of them here) :

((Figure 6))

Both models **2** and **3** contain a formally high-spin $Fe^{3+}$ ion (*ie* with empty minority spin orbitals). We found however a significant fraction (~ 0.6 for model **2**, ~ 0.8 for model **3**) of spin spread over two occupied (minority spin) orbitals. The first of these molecular orbitals ($\phi_{xy}$) results from the interaction of the Fe $d_{xy}$ atomic orbital (~ 30 - 40%) with a peroxo $Op_x$ - $Op_x$ linear combination (~ 50%) and is similar for **2** and **3**. The second one ($\phi_{yz}$) which results from the interaction of Fe $d_{yz}$ (~ 30 - 40%) with a peroxo $Op_z$ - $Op_z$ linear combination differs between both models. In the case of model **3**, $\phi_{yz}$ contains a large peroxo contribution (~ 40%). By contrast, in model **2**, $\phi_{yz}$ contains a large sulfur $p_y$ contribution (~ 30%) and the peroxo contribution is reduced (~20 %). The HOMO-LUMO gaps are 0.8 eV for model **2** and 0.7 eV for model **3**. We have verified that the Slater transition energies (see the Experimental Section) do not differ much from the d (occupied) to d (empty) transitions derived directly from the ground state electronic structures (such is not the case for formal $Fe^{2+}$ ions, in our experience). As the deviations from the free electron value $g_e$ = 2.0023 are very small indeed for a $Fe^{3+}$ ion (because of the rather large d-d gaps involved), and in spite of the slight "$Fe^{2+}$" character resulting from the Fe-($O_2$) interaction, the computation of [g] and [D] tensors is tentative only. The numerical results and the ZFS parameters D and E/D, as well as the largest (in magnitude) principal values of the quadrupole tensors for **2** and **3** are reported in Table 2 :

((Table 2))

Within such limits, both [g] tensors are very similar for **2** and **3**, as far as the principal values (cf Table 3) and axes (not shown) are concerned. Models **2** and **3** exhibit small D values (0.50 and 0.56 cm$^{-1}$, respectively) that are consistent with a high-spin Fe$^{3+}$ ion. Both quadrupole splittings values for **2** and **3** are found negative ($\Delta E_Q$ = - 1.04 mm/s and - 1.40 mm/s, respectively). The computed quadrupole tensors are rather rhombic in the case of **2** ($\eta$ = 0.65) and axial in the case of **3** ($\eta$ = 0.13). This difference originates directly from the relative weights of $d_{xy}$ and $d_{yz}$ in $\phi_{xy}$ and $\phi_{yz}$ due to the presence of the thiolate ligand in **2**.

*(c) Isomer shift prediction.* Unlike the quadrupole splitting, the direct calculation of the isomer shift is not straightforward (*45*). It is possible to estimate it by establishing a correlation between experimental isomer shifts and theoretical electron densities at the iron nucleus $\rho(0)$, as previously done in the literature (*45-49*). Since the computed electron densities depend on the used DFT exchange-correlation potential, we built a new correlation following this procedure. The mononuclear Fe$^{3+}$ complexes and biomolecules used for establishing the linear correlation are shown in Table 3 :

((Table 3))

The electron densities at the $^{57}$Fe nuclei were plotted *vs* the experimental isomer shifts, and the resulting graph was subjected to a very satisfactory linear regression analysis, as shown in Figure 7 :

((Figure 7))

Linear regression analyses of these data gave $\delta$(mm/s) = 7.34 – 0.270$\rho(0)$ (r = 0.992) for the Fe$^{3+}$ series. By varying the Fe-S distance from 2.46 Å to 2.30 Å in the model of oxidized center II in SOR from *D. Desulfuricans* (model **2**, key : 2 in Figure 7), the associated point in Figure 7 remains close to the Fe$^{3+}$ linear correlation (data not shown). When using the Fe$^{3+}$ correlation of Figure 7 to estimate the isomer shift values for **2** and **3**, three main features are noticeable : i) the calculated value for model **2** compares favorably (0.56 mm/s *vs* 0.54 (1) mm/s) with the experimental value of the peroxo-Fe$^{3+}$ species in E47A SOR from *D. baarsii*, ii) replacing an axial nitrogen (model **3**) by an axial thiolate (model **2**) ligand for the Fe$^{3+}$ ion brings about a significant decrease (- 0.07 mm/s) of the isomer shift value, in line with that observed when going from **6** to **4** and **5** (see above)  iii) the isomer shift value of 0.63 mm/s estimated for **3** is consistent with experimental data available in the literature for similar species (*17,19*).

# DISCUSSION

Our initial experiments on the peroxo-$Fe^{3+}$ species of E47A SOR from *D. baarsii* reacted with $H_2O_2$ led us to propose, on the basis of RR, that it is a side-on peroxo-$Fe^{3+}$ species (*16*). Such species possess specific signatures in Mössbauer spectroscopy which distinguish them clearly from the alternative end-on hydroperoxo form (see below). Therefore, to further characterize this peroxo-$Fe^{3+}$ species, we have obtained in the present study a complete set of Mössbauer and spin-Hamiltonian parameters.

Recent spectroscopic and magnetic studies have established that the vast majority of mononuclear hydroperoxo-$Fe^{3+}$ complexes with only N/O ligands exhibit: (i) a low-spin ground state S = ½ (*52*), (ii) isomer shift values within the range 0.16 - 0.19 mm/s (*17,19,42,53*), (iii) ν(Fe-O) vibrations in the range 615 - 645 $cm^{-1}$ and (iv) ν(O-O) vibrations in the range 780 - 810 $cm^{-1}$ with a deuterium isotopic shift (*17,52*). A single hydroperoxo-$Fe^{3+}$ compound departs from this behavior, being high-spin with a ν(O-O) vibration at 830 $cm^{-1}$ (*54*). In contrast, mononuclear side-on peroxo-$Fe^{3+}$ complexes of N/O ligands exhibit: (i) a high-spin ground state S = 5/2 (*52*), (ii) isomer shift values within the range 0.61 - 0.65 mm/s (*17,19,41*), (iii) ν(Fe-O) vibrations in the range 450 - 500 $cm^{-1}$ and (iv) ν(O-O) vibrations in the range 815 - 830 $cm^{-1}$ insensitive to hydrogen/deuterium exchange (*17,52*).

The parameters obtained in this work for the peroxo-iron species in E47A SOR from *D. baarsii* are consistent with a high-spin ground state S = 5/2. In addition, it appears that both the isomer shift and the asymmetry parameter values of the peroxo species are higher than those determined for the center II oxidized with $K_2IrCl_6$ (Table 1). This observation is in agreement with the presence of the peroxo ligand. However, the value of the isomer shift (0.54 (1) mm/s) appears outside the range observed for complexes of N/O ligands (see above). We reasoned that this might be caused by the presence of the axial cysteinate ligand in SOR which is absent in the side-on peroxo-$Fe^{3+}$ complexes reported so far. This hypothesis was validated by several lines of evidence (i) the experimental observation that replacing a carboxylate ligand by a thiolate ligand in $Fe^{2+}$ model complexes lowers the isomer shift value by *ca* 0.1 mm/s, (ii) the results of DFT calculations showing a similar decrease of the isomer shift value ($\Delta\delta_{/Fe}$ = - 0.07 mm/s) when an axial $NH_3$ ligand of the $Fe^{3+}$ ion in model **3** is replaced by a methylthiolate ligand in model **2**, (iii) the axial ZFS, the quadrupole splitting and the asymmetry parameter values that have been evaluated by DFT calculations for the $\eta^2$- $O_2$ $Fe^{3+}$ model **2** (D = 0.5 cm$^{-1}$, $\Delta E_Q$ = -1.04 mm/s and $\eta$ = 0.65) match reasonably well the experimental values obtained for the peroxo-$Fe^{3+}$ species in E47A SOR (D = 0.8 (2) cm$^{-1}$, $\Delta E_Q$ = -0.80 (5) mm/s and $\eta$ = 0.60 (5)). It is worth noting that the calculated value of the isomer shift of **3** falls within the experimental range expected for $\eta^2$-$O_2$ $Fe^{3+}$ complexes with nitrogen ligands (see above), what supports the validity of the calculations. Therefore, the isomer shift value for a high-spin $\eta^2$-$O_2$ $Fe^{3+}$ complex with a thiolate ligand can be estimated to be in the 0.54 – 0.58 mm/s range. The isomer shift value determined here for the E47A SOR peroxo species fits this range.

It follows then that all spectroscopic features of the E47A SOR peroxo species from *D. baarsii* are consistent with a side-on peroxo-$Fe^{3+}$ formulation and that a hydroperoxo-$Fe^{3+}$ formulation must be rejected. This would imply an hepta-coordination of the $Fe^{3+}$ ion in the peroxo complex. Indeed, the RR band at 743 cm$^{-1}$, which corresponds to a C-S stretching mode of the CysS-$Fe^{3+}$ active site, is still observed in the peroxo species. This indicates that the cysteinate ligand is still coordinated to the iron center. In addition, there is no evidence for a His(N)-Fe rupture upon addition of an extra ligand in SOR from *P. furiosus* (*7*). In this respect, it is of interest that a seven-coordinated $\eta^2$-$O_2$ $Fe^{3+}$complex has been very recently reported in the literature (*17*).

## CONCLUSION

The E47A mutation in SOR from *D. baarsii* allows the stabilization of a peroxo-iron species in the second time scale when reacted with $H_2O_2$. We were able to prepare a sample of this peroxo species by manual freezing in liquid nitrogen and to study it by spectroscopic methods. All the experimental and theoretical evidences presented here show that this peroxo complex is a high-spin side-on peroxo-$Fe^{3+}$ species and clearly rule out a hydroperoxo-$Fe^{3+}$ species. Therefore, a mutant of SOR from *D. baarsii* (E47A mutant) is able to accommodate a side-on peroxo complex without severe disruption of its protein matrix (*18*). It is of interest that very recently a similar side-on peroxo-$Fe^{3+}$ adduct has been crystallographically characterized for the first time in a protein in the case of naphthalene dioxygenase (*35*). The kinetic and UV-visible analyses of the peroxo intermediates detected in the reaction of superoxide with the SOR of various origins (*11-14*) suggest that the same species is formed initially in all cases. Its decomposition, probably following a protonation process, varies with the origin of the enzyme. Therefore, a $\eta^2$-$O_2$ $Fe^{3+}$ species may be considered as a candidate for an intermediate involved in the catalytic cycle of SORs. Experiments are presently underway in our laboratories to further characterize these intermediates.

## ACKNOWLEDGMENTS

Dr. C. Philouze (Université Joseph Fourier), Dr T. Kent (Web Research) and Prof. L. Noodleman (Scripps Research Institute) are gratefully acknowledged for useful discussions. N. Genand-Riondet (CEA/Saclay) is gratefully thanked for assistance in the in-field Mössbauer experiments.

## SUPPORTING INFORMATION AVAILABLE

Zero-field Mössbauer spectra and high-field and high-temperature Mössbauer spectrum of $^{57}$Fe SOR E47A from *D. baarsii* after reduction by sodium dithionite (Figure S1), high-field Mössbauer spectra of $^{57}$Fe SOR E47A from *D. baarsii* treated with sodium dithionite (Figure S2), Resonance Raman spectrum of $^{57}$Fe SOR E47A treated with $H_2O_2$ (Figure S3), zero-field Mösbauer spectra (T = 4.2 K) of complexes **4**, **5** and **6** (Figure S4), zero-field Mössbauer parameters of complexes **4**, **5** and **6** (Table S5), coordinates in Å for the model of oxidized center II in SOR, the models **2** and **3**, and the model of reduced center II in SOR used for DFT calculations (Tables S6–S9). This material is available free of charge via the Internet at http://pubs.acs.org.

# TABLES

**Table 1.** Mössbauer parameters for center I and center II in SOR E47A from *D. baarsii* at 4.2 K (a) for the as-isolated enzyme, (b) after oxidation of the enzyme by $K_2IrCl_6$ and (c) after oxidation of the enzyme by $H_2O_2$. (d) Set to the value determined by magnetization measurements.

**Table 2.** Eigenvalues of the computed [g], [D] and quadrupole tensors ($\eta$ = asymmetry parameter) and calculated theoretical electron densities $\rho(0)$ at the iron nucleus for models **2** and **3**.

**Table 3.** Mononuclear iron complexes and biomolecules (column 1) used for establishing the linear correlation between calculated theoretical electron densities at the iron nucleus (column 3) and experimentally measured isomer shifts (column 4).

**Table 1.**

| Iron center | Center I ($Fe^{2+}$) in $S_2O_4^{2-}$ reduced E47A SOR | Center I ($Fe^{3+}$) in as-isolated, $IrCl_6^-$ and $H_2O_2$ oxidized E47A SOR | Center II ($Fe^{2+}$) in $S_2O_4^{2-}$ reduced and as-isolated E47A SOR | Center II ($Fe^{3+}$) in $IrCl_6^-$ oxidized E47A SOR | Center II ($Fe^{3+}$) in $H_2O_2$ oxidized E47A SOR |
|---|---|---|---|---|---|
| $D$ (cm$^{-1}$) | $-6.0^a$ | $2.2^a$ | $5.2^c$ | $-1.3\,(2)$ | $0.8\,(2)$ |
| $E/D$ | $0.19^a$ | $0.08^a$ | $0.28\,(3)^c$ | $0.33^e$ | $0.33^f$ |
| $g_x$ | $2.08^b$ | $2.0$ | $2.08^d$ | $2.0$ | $2.0$ |
| $g_y$ | $2.02^b$ | $2.0$ | $2.15^d$ | $2.0$ | $2.0$ |
| $g_z$ | $2.20^b$ | $2.0$ | $2.00^d$ | $2.0$ | $2.0$ |
| $A_x/g_n\beta_n$ (T) | $-20.4\,(2)$ | $-15.4\,(4)$ | $-23.4\,(2)$ | $-20.1\,(5)$ | $-21.6\,(5)$ |
| $A_y/g_n\beta_n$ (T) | $-20.4\,(2)$ | $-15.4\,(4)$ | $-7.4\,(2)$ | $-20.0\,(8)$ | $-20.5\,(3)$ |
| $A_z/g_n\beta_n$ (T) | $-6.5\,(2)$ | $-15.4\,(4)$ | $-8.1\,(2)$ | $-21.0\,(5)$ | $-21.0\,(5)$ |
| $\Delta E_Q$ (mm/s) | $+3.37\,(3)$ | $-0.79\,(4)$ | $+2.82\,(3)$ | $-0.53\,(5)$ | $-0.80\,(5)$ |
| $\eta$ | $0.35\,(5)$ | $1.0\,(2)$ | $0.40\,(5)$ | $0.00\,(5)$ | $0.60\,(5)$ |
| $\delta_{/Fe}$ (mm/s) | $0.69\,(2)$ | $0.29\,(1)$ | $1.06\,(1)$ | $0.47\,(1)$ | $0.54\,(1)$ |
| Euler angles (°) | $0\,;\,10\,;\,0^a$ | $0\,;\,90\,;\,0^a$ | $78\,;\,40\,;\,0$ | $0\,;\,90\,;\,90^g$ | $17\,;\,80\,;\,60$ |
| FWHM$^h$ | $0.30$ | $0.35$ | $0.33$ | $0.35$ | $0.35$ |

(a) reference 38, (b) determined from the relations $g_x = g_z - 2(D - E)/\lambda$, $g_y = g_z - 2(D + E)/\lambda$ and $g_z = 2.20$, where $\lambda = -80$ cm$^{-1}$ is the spin orbit coupling constant, (c) from EPR and magnetization measurements (this work), (d) determined from the relations $g_x = g_z - 2(D - E)/\lambda$, $g_y = g_z - 2(D+E)/\lambda$ and $g_z = 2.0$, where $\lambda = -100$ cm$^{-1}$ is the spin orbit coupling constant, (e) reference 16, (f) from EPR measurements (this work), (g) reference 8, (h) FWHM = Full Width at Half Maximum (in mm/s).

**Table 2.**

|  | Model **2** | Model **3** |
|---|---|---|
| $g_1$ | 2.010 | 2.012 |
| $g_2$ | 2.009 | 2.011 |
| $g_3$ | 2.006 | 2.006 |
| D (cm$^{-1}$) | 0.50 | 0.56 |
| E/D | 0.07 | 0.06 |
| $\Delta E_Q$ (mm/s) | -1.04 | -1.40 |
| $\eta$ | 0.65 | 0.13 |
| $\rho(0)$ (au$^{-3}$) | 25.0766 | 24.8595 |
| $\delta_{/Fe}$ (mm/s) | 0.55 | 0.60 |

**Table 3.**

| Compound | ref. | $\rho(o)$ (au$^{-3}$) | $\delta_{/Fe}$ (mm/s) | ref. |
|---|---|---|---|---|
| Model of [Fe(edta)(O$_2$)]$^{3-}$ | 33 | 24.791 | 0.65 | 42 |
| Oxidized center II in SOR | PDB - 1DQI | 25.270 | 0.50 | 8 |
| [FeCl$_4$]$^{1-}$ | 44 | 25.948 | 0.36 | 47 |
| [Fe(SR)$_4$]$^{1-}$ | PDB – 1FHH | 26.136 | 0.24 | 50 |
| [Fe(NO$_2$)(Py)(TpivPP)] | ICSD - SOBZUE | 26.346 | 0.26 | 51 |
| [Fe(CN)$_6$]$^{3-}$ | ICSD - 200200 | 27.258 | -0.03 | this work |

## SCHEME

**Scheme 1.**

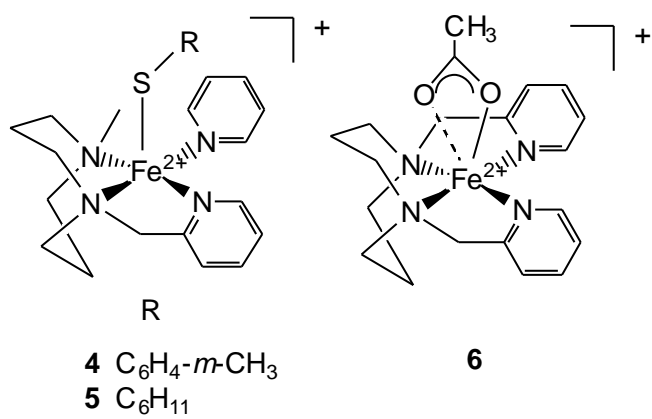

**4** $C_6H_4$-*m*-$CH_3$
**5** $C_6H_{11}$

**6**

**Scheme 2.**

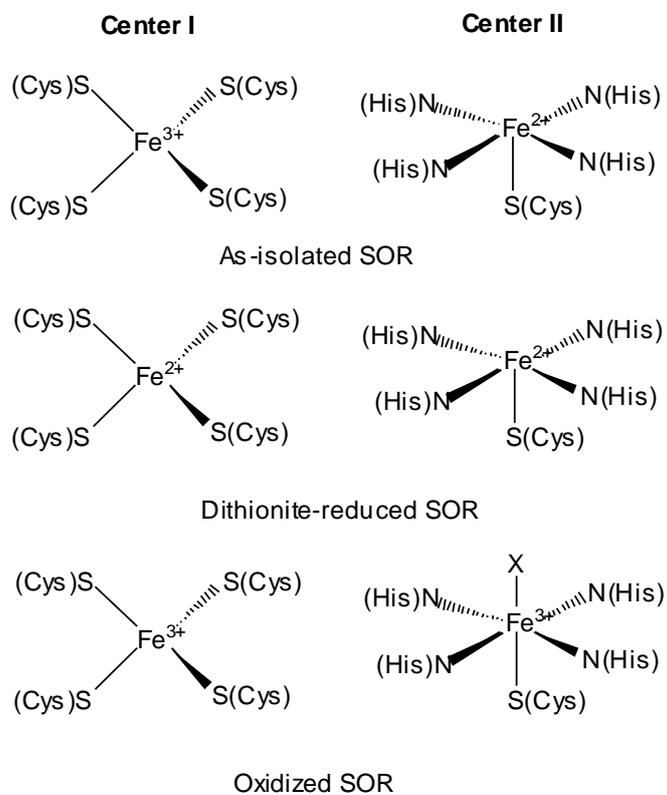

As-isolated SOR

Dithionite-reduced SOR

Oxidized SOR

# FIGURES

**Figure 1.** The three theoretical models of the SOR active site that were investigated by DFT calculations in this study.

**Figure 2.** βB/kT ratio dependence of the molar magnetization of the as-isolated SOR *from D. baarsii* (5.1 mM in 50 mM Tris/HCl buffer in $D_2O$, pH = 7.6). The isofield experiments were performed at 0.5, 1.0, 2.0, 3.0, 4.0 and 5.0 T. The lines through the points correspond to the best fit obtained with 50% contribution of center I (g = 2.0, D = 2.2 $cm^{-1}$ and E/D = 0.08) and 50 % contribution of center II (g = 2.15, |D| = 5.2 $cm^{-1}$ and E/D = 0.28 ± 0.05).

**Figure 3.** Mössbauer spectrum of as-isolated $^{57}Fe$ SOR E47A from *D. baarsii* (4.5 mM in 50 mM Tris/HCl buffer, pH = 7.6) at 4.2 K, in a magnetic field of 50 mT applied parallel to the γ-beam. The solid curves show the contribution of each iron site ((a) : $Fe^{2+}$ from center II, (b) $Fe^{3+}$ from center I).

**Figure 4.** Mössbauer spectra of $^{57}Fe$ SOR E47A from *D. baarsii* (1.5 mM in 50 mM Tris/HCl buffer, pH = 7.6) at 4.2 K, treated with 3 equivalents of $K_2IrCl_6$. The experimental spectra taken at 4.2 K in a magnetic field of (A) 50 mT, (B) 1.5 T, (C) 3.0 T, (D) 5.5 T applied parallel to the γ-beam were fitted (solid curves) with the parameters set of Table 1. The solid curves above the experimental spectra show the contribution of each iron site ((a) : $Fe^{3+}$ from center I, (b) : $Fe^{3+}$ from oxidized center II, (c) $Fe^{2+}$ from center II).

**Figure 5.** Mössbauer spectra of $^{57}Fe$ SOR E47A from *D. baarsii* (1.5 mM in 100 mM Tris/HCl buffer, pH = 8.4) at 4.2 K, treated with 6 equivalents of $H_2O_2$ and immediately frozen in liquid nitrogen. The experimental spectra taken at 4.2 K in a magnetic field of (A) 50 mT, (B) 1.5 T, (C) 3.0 T, (D) 5.5 T applied perpendicular to the γ-beam were fitted (solid curves) with the parameters set of Table 1. The solid curves above the experimental spectra show the contribution of each iron site ((a) : $Fe^{3+}$ from center I, (b) : $Fe^{3+}$ from oxidized center II, (c) $Fe^{2+}$ from center II).

**Figure 6.** Simplified description of the (DFT spin unrestricted) electronic structures of models **2** and **3** (only some of the spin minority β orbitals are represented). The $\phi_{xy}$ molecular orbital results from the interaction of the Fe $d_{xy}$ atomic orbital (~ 30 - 40%) with a peroxo $Op_x$ -

Op$_x$ linear combination (~ 50%), and the $\phi_{yz}$ molecular orbital results from the interaction of Fe d$_{yz}$ (~ 30 - 40%) with a peroxo Op$_z$ – Op$_z$ linear combination. "a" refers to a combination of Sp$_y$ (50%) and peroxo Op$_z$ (30%) orbitals. "b" refers to a combination of imidazole C and Np$_y$ orbitals. The z$^2$ and (x$^2$ – y$^2$) molecular orbitals contain a small fraction of the (x$^2$ – y$^2$) and z$^2$ orbitals, respectively. The meaning of "HOMO" and "LUMO" is restricted here within the minority spin orbitals

**Figure 7.** Calibration of the «VBP» method for prediction of $^{57}$Fe isomer shifts. The calculated electron density at the iron nucleus is plotted *vs* the experimentally determined isomer shifts for a series of mononuclear iron complexes and biomolecules (see Table 1 for details). The solid lines correspond to the linear correlation analyses of the Fe$^{3+}$ data. Key : (1) model of [Fe(EDTA)(O$_2$)]$^{3-}$; (2) model of oxidized center II in SOR from *D. Desulfuricans*; (3) [FeCl$_4$]$^{1-}$; (4) [Fe(SR)$_4$]$^{1-}$; (5) [Fe(NO$_2$)(Py)(TpivPP)]; (6) [Fe(CN)$_6$]$^{3-}$.

**Figure 1.**

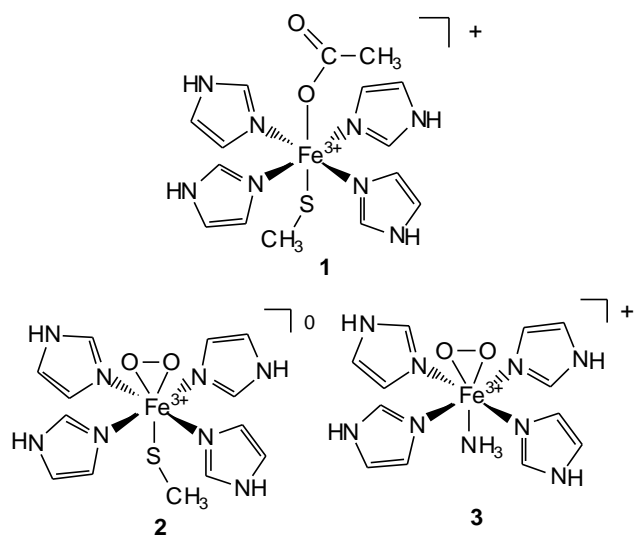

**Figure 2.**

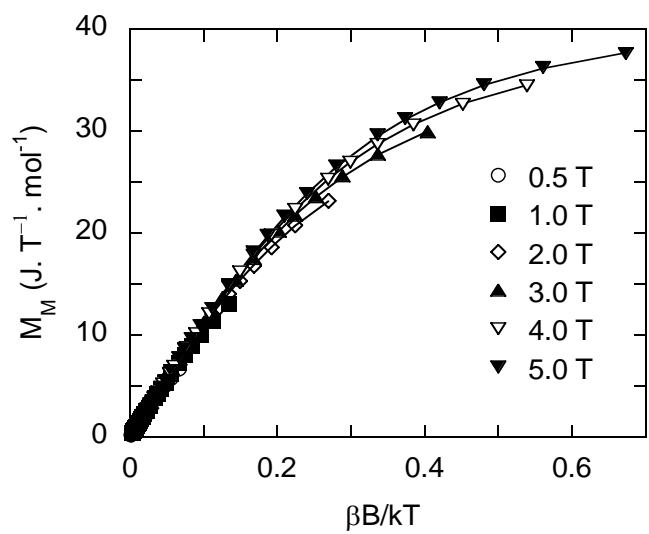

**Figure 3.**

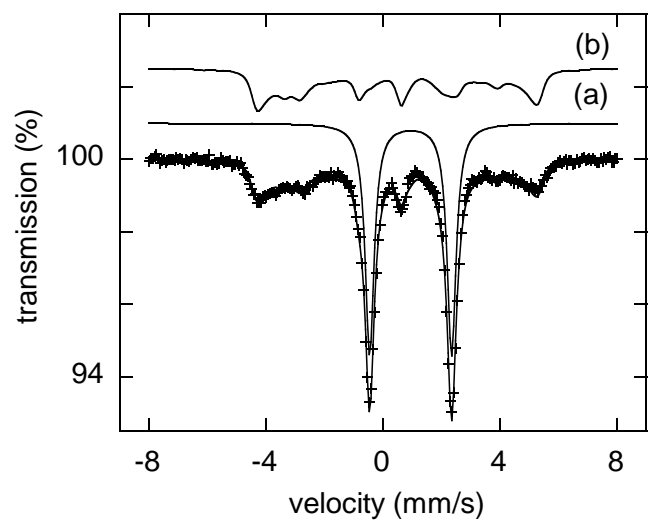

**Figure 4.**

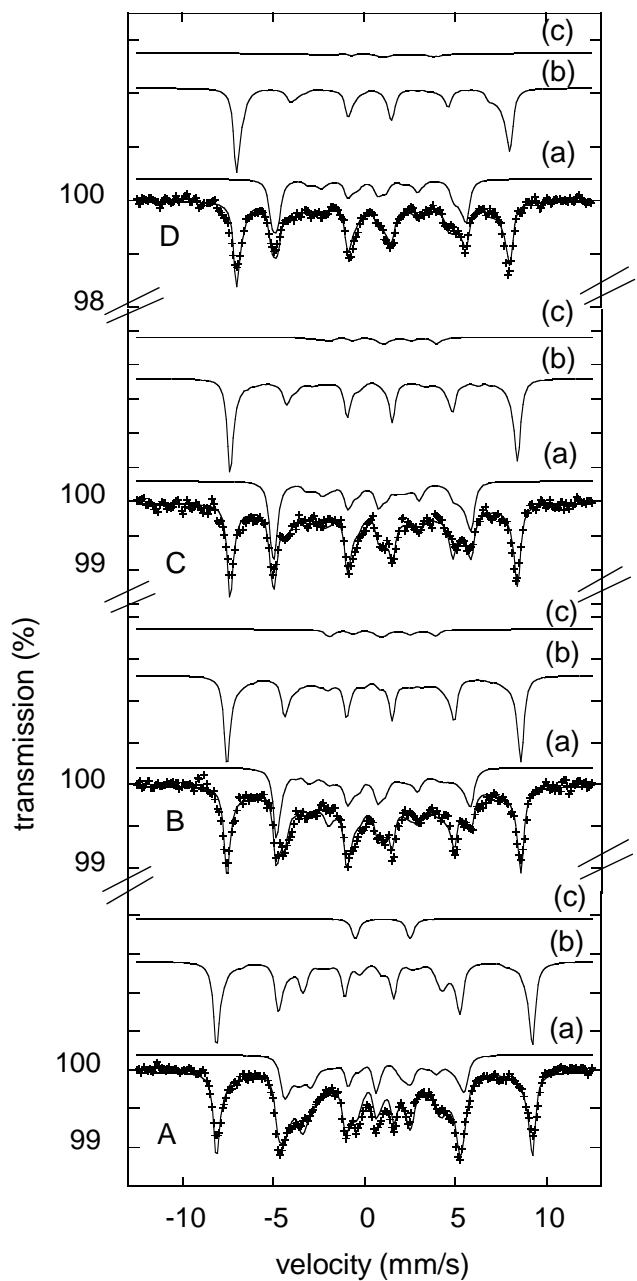

**Figure 5.**

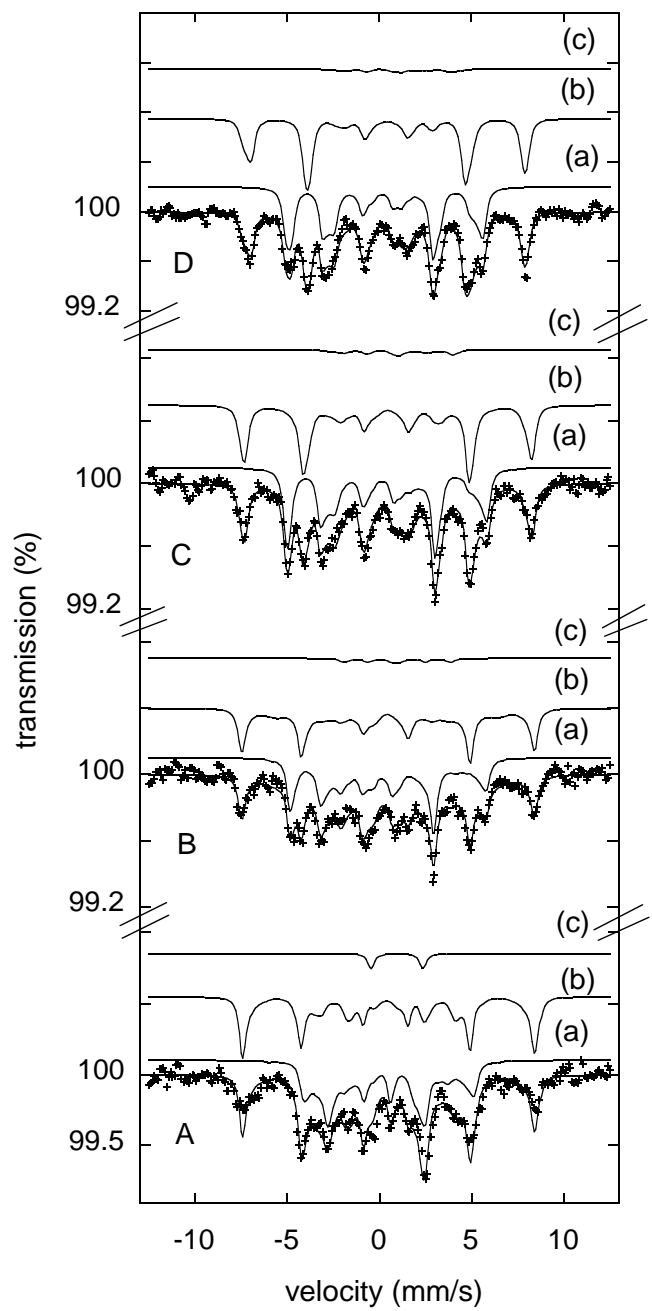

**Figure 6.**

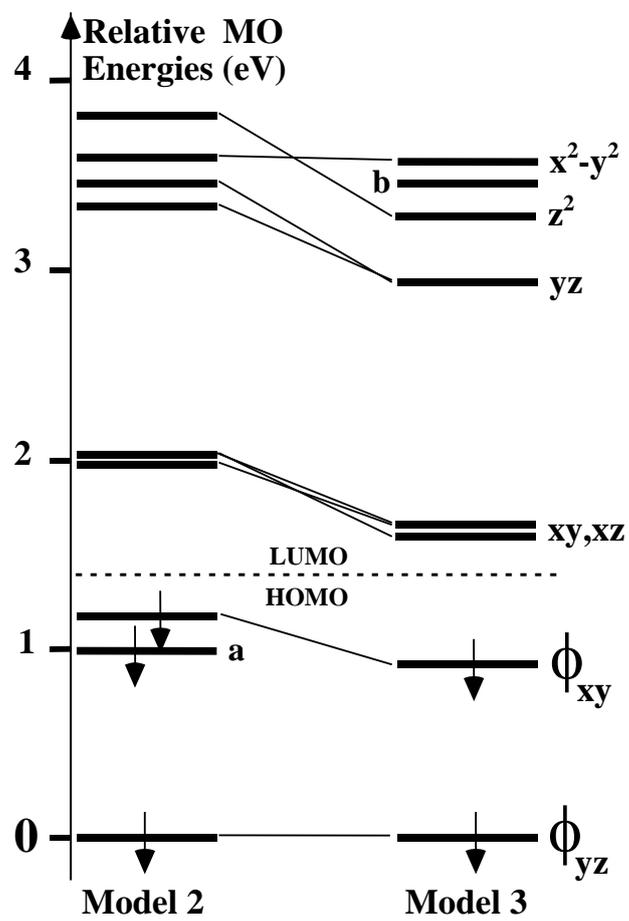

**Figure 7.**

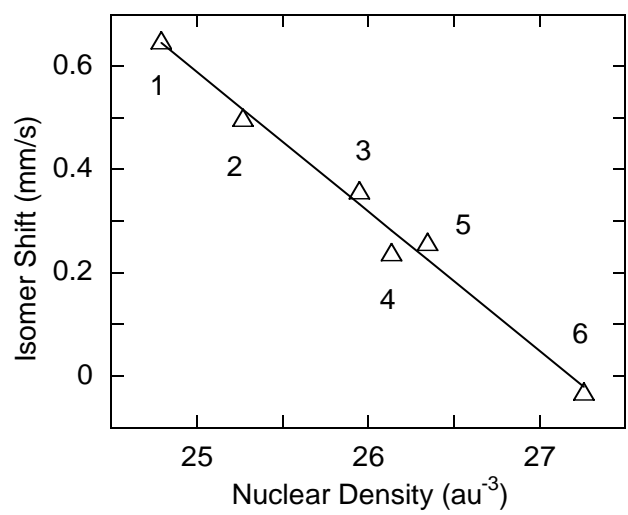

343**For Table of Contents Use Only**

Mössbauer Characterization of an Unusual High-spin Side-on Peroxo-$Fe^{3+}$ Species in the Active Site of Superoxide Reductase from *Desulfoarculus baarsii*. Density Functional Calculations on Related Models

Olivier Horner, Jean-Marie Mouesca, Jean-Louis Oddou, Claudine Jeandey, Vincent Nivière, Tony A. Mattioli, Christelle Mathé, Marc Fontecave, Pascale Maldivi, Pierre Bonville, Jason A. Halfen, and Jean-Marc Latour

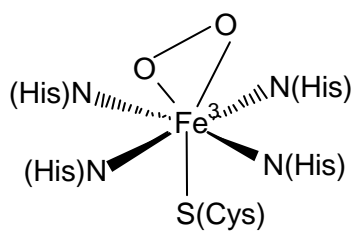

Center II in SOR from *D. baarsii* reacted with $H_2O_2$